\documentstyle[preprint,aps,prl,overcite]{revtex}
\input epsf

\begin{document}

\title{
 {\sc Parametric Resonance in an Expanding Universe
        }} 

\author
{Ivaylo Zlatev, Greg Huey, and Paul J. Steinhardt}

\address{Department of Physics and Astronomy,\\
University of Pennsylvania \\ Philadelphia, PA 19104 USA}

\maketitle

\begin{abstract}

Parametric resonance has been discussed as a mechanism for copious particle
production following inflation.
Here we present a simple and intuitive calculational method for estimating 
the efficiency of parametric amplification as a function of 
parameters.   This is important for determining whether 
resonant amplification plays an important role in the reheating 
process.  
We find that significant amplification occurs only for a limited range of 
couplings and interactions.

\end{abstract}

\newpage

\narrowtext

Reheating after the end of inflation is the process during which almost all
elementary particles in the universe are presumed to have been created.
In versions of inflationary cosmology
in which inflation ends by slowly rolling towards
the minimum of the inflaton potential,\cite{Linde82}$^,$\cite{Alb82}
reheating occurs as the inflaton
field $\phi$ descends to the minimum and begins to oscillate 
rapidly about it.
The oscillating inflaton field $\phi$
decays to the  various matter fields to which it is 
coupled.\cite{Alb82a,Dol82,Abb82}
In the typical case, the decay of the inflaton can be represented
by adding to the equation-of-motion for $\phi$
a phenomenological term $\Gamma \dot{\phi}$ where $\Gamma$ is  the effective
decay rate of $\phi$ particles;
a corresponding term is added to the equation describing
radiation to account for the transfer 
of energy from $\phi$ to radiation.\cite{Alb82a}
This treatment assumes the $\phi$ particles composing the oscillating
$\phi$-field decay incoherently.

Recently, there has been interest in a two-stage reheating scenario which
begins with the coherent transfer
of energy from the oscillating
 inflaton $\phi$ to a second field (or fields) $X$ through
parametric 
resonance.\cite{Kofman94,Boyanovsky95,Shtanov95,Kofman96,Kolb96,Klebnikov96,Klebnikov97,Fujisaki96,Kofman97,Greene97,Klebnikov96a,Klebnikov97a,B.Greene97,Allahverdi97,Prokopec97,Riotto97}
Later, the $X$-particles decay to a rapidly thermalizing distribution of
ordinary matter, completing the reheating process.
During the first stage, which has been named
{\it preheating}, the density of $X$ particles undergoes a period of
exponential increase.
The resonant transfer of energy is important to inflationary
cosmology if it is  ``efficient": if it
 results in  amplified production of $X$ particles
 by many orders of magnitude compared to estimates
based on the decay rate ($\Gamma$) of an individual inflaton particle
such that a large fraction of 
the inflaton energy density is  converted
to $X$ particles.
In addition to producing a higher reheating temperature,
 preheating enables the coherent transfer of  inflaton energy
 to particles $X$ which are more massive
than $\phi$ itself, whereas 
only particles less massive than $\phi$
are produced  in incoherent decay and the density of very 
massive $X$-particles
is exponentially suppressed during  subsequent thermalization.
This makes possible  models of baryogenesis which
rely on there being a high density of  particles
more massive than $\phi$ following inflation which subsequently
undergo baryon-asymmetric decays.\cite{Kolb96}

The goal of this paper is to present a straightforward and intuitive method for
determining the efficiency of parametric resonance for a wide range of
model couplings and interactions.  We present results for  a simple
model in which
the $X$-particle is massive and has quartic self-interactions.
The method predicts and makes intuitively clear why
resonant particle production is efficient only for a narrow range of 
masses and couplings. 
To check the prediction,
we have developed precise numerical codes similar to those
of Tkachev and Klebnikov\cite{Klebnikov96,Klebnikov97} and, as shown in the paper,
we have found that the estimates of particle production compare very well.
Where there is overlap, the results are also consistent with the analysis
by Kofman {\it et al.}.\cite{Kofman97,Greene97}
With this verification, 
this approach can now be applied to general inflaton models to determine if 
resonant particle production is a significant feature of the reheating process.

In our method, resonance is described by trajectories through stability/instability
regions of the Mathieu equation.  This notion has been alluded to in previous
discussions;\cite{Kofman94,Klebnikov97,Kofman97,B.Greene97} 
here we show that it can be developed into a 
reliable, predictive tool.  Quantitative estimates are simply obtained from 
calculable properties of the Mathieu equation and the trajectories. 
The estimates are reliable except for  very large couplings such
that other nonlinear effects --- stochastic resonance, backreaction,  
rescattering and chaotic instability --- become important.  
Although this large coupling regime has been the focus of many 
interesting studies,\cite{Kofman94,Klebnikov97,Kofman97,Klebnikov96a,B.Greene97}
it  lies far into the ``efficient preheating"
regime;  as we will demonstrate, we do not need to consider these complications 
for our purpose of finding
the boundary between inefficient and efficient preheating.

The general model which we  consider  is described by
the  Lagrangian density:
\begin{eqnarray}
        {\cal L} = \frac{1}{2}(\partial_{\mu} \phi)^2 +
\frac{1}{2}(\partial_{\mu} X)^2
                - \frac{1}{2}g^2 \phi^2 X^2
                -\frac{1}{2}m_{\phi}^2\phi^2 - \frac{1}{2}m_{x}^2X^2
                - V(X).
\end{eqnarray}
Here
$V(X)$ represents self-couplings of the $X$-field of cubic order or higher.
If we assume that $\phi$ is spatially uniform, we can derive a set
of coupled equations for $\phi$ and the Fourier modes , $X_k$.
The equations for  $X_k$ decouple from one another in the Hartree (mean
field) approximation.
 In terms of ``comoving"
fields, $\Phi\equiv a^{3/2}\phi$ and $\chi_k \equiv a^{3/2} X_k$,
the linearized equations are:
\begin{eqnarray}
  \label{eom1}      &&\ddot{\Phi} + \left[m_{\phi}^2
                + \frac{g^2 \langle\chi^2\rangle}{a^3} +6\pi G p\right]\Phi =0
\\
  \label{eom2}      &&\ddot{\chi}_k
 +\left [\omega_k^2 +  \frac{g^2 \Phi^2}{a^3}\right] \chi_k= 0,
\end{eqnarray}
where
\begin{eqnarray}
\omega_k^2 \equiv \frac{k^2}{a^2}+m_x^2 + f(\chi) +
6\pi G p \\
p= -\frac{1}{4 \pi G}(\frac{\ddot{a}}{a} +\frac{1}{2}
                \frac{\dot{a}^2}{a^2}) \\
f(\chi) \equiv \langle \frac{ a^{3/2} V'(\chi/a^{3/2})}{\chi}
\rangle
\end{eqnarray}
where $p$ is the pressure,  $a$ is
the scale factor, $G$ is
Newton's constant,
and $\langle \ldots \rangle$ represents the spatial average.
Henceforth, we choose units of energy, mass and time
where $m_{\phi}=1$.

The pressure is negligible when the universe is dominated by
inflaton oscillations.
Furthermore, until one reaches the condition
where   $g^2 \chi^2/a^3  \gtrsim  m_{\phi}^2$,
 the backreaction\cite{Kofman94,Klebnikov97,Kofman97,Klebnikov96a,B.Greene97} 
 on $\Phi$ due to $X$ and rescattering are   negligible, too.
This 
condition is not satisfied as long as the energy density in $X$-particles,
$\rho_x = \omega_x^2 \langle \chi^2/a^3 \rangle= {\cal O}(1)
(g^2 \Phi^2 + m_x^2) \langle \chi^2/a^3 \rangle$,  does not exceed the
inflaton energy density, $m_{\phi}^2 \Phi^2$. But, this is precisely
our regime 
of interest since we wish to determine the range of 
parameters for which resonant amplification is inefficient to 
barely efficient ($\rho_x \le \rho_{\phi}$).
Hence, for our
purpose of finding the boundary between
the inefficient and efficient regimes, backreaction and rescattering can be 
safely neglected.  Of course, for previous studies in the highly efficient,
strongly coupled regime, backreaction and rescattering 
cannot be 
ignored\cite{Kofman94,Klebnikov97,Kofman97,Klebnikov96a,B.Greene97}.

Hence, to good approximation, $\Phi$
behaves as a simple harmonic oscillator, $\Phi=
\Phi_0 \, {\rm cos}\, t$, where time
$t$ is measured in units of $m_{\phi}=1$ beginning from initial
time $t_0 \approx m_{\phi}^{-1}$ when the inflaton begins to
oscillate.
Then, the equation-of-motion for $\chi$ can be written:
\begin{eqnarray} \label{matt}
 &&\frac{\partial ^2 \chi_k}{\partial t^2} +[A+2q\, {\rm cos} (2t)]\chi_k = 0
\end{eqnarray}
where
\begin{eqnarray}
        &&q= \frac{g^2 \Phi_0^2}{4 a^3} \nonumber \\
        &&A=  \omega_k^2+  \frac{g^2 \Phi_0^2}{2 a^3} = \omega_k^2 +2 q
\end{eqnarray}

In the limit of a static universe, $a \rightarrow$~constant,
Eq.~(\ref{matt}) is precisely a Mathieu equation.\cite{McLachlan}
The $(q,\,A)$ plane
is separated into stable regions where the amplitude of $\chi$
stays constant and
unstable regions where the amplitude increases exponentially, $\chi \propto
e^{\nu t}$,
where the critical exponent 
$\nu$  (the  imaginary part of
the Mathieu characteristic exponent)  depends on $q$ and $A$.
See Figure 1.
For fixed $A$, $\nu$ grows and the instability strips
broaden as   $(q,\, A)$
moves horizontally towards the right.
The dotted line $A=2q$
roughly divides the strips into their narrow and broad regimes.
Each mode $k$ corresponds to some 
fixed   $(q,\,A)$; if this point lies in an unstable regime, the 
mode is exponentially amplified.  
Increasing $k$ corresponds to increasing
$A$ leaving $q$ unchanged, so the modes lie along a ray pointing 
vertically upwards beginning with the $k=0$ mode.

Important differences arise  in an expanding universe.
   First and foremost, $A$
and $q$ are functions of the scale factor $a$ and, therefore, vary with time.
In our treatment, we propose to use the Mathieu equation
as a guide but
take account of the time-evolution in the $(q,\, A)$ plane.
For simplicity, we first consider $f(\chi)=0$.
As shown in Figure 1, $(q =q_0/ a^{-3}, \, A= 2q + (k/a)^2+m_x^2)$
is no longer a fixed point in an expanding universe, $a=a(t)$. 
  Rather,
each $k$ corresponds to a trajectory in the $(q,\,A)$ plane from upper right,
$(q_0,\, A_0)$,
to lower left, $(q,\, A) \rightarrow (0,m_x^2)$.
For $k>0$, 
 the starting point is shifted vertically upwards
from $(A_0=2 q_0 + m_x^2)$ by $k^2$.  Then,  the mode is red shifted and
the  trajectory  curves towards
the $A=2q+m_x^2$ line as $q =g^2 \Phi_0^2/4a^3\rightarrow 0$.

Resonant amplification occurs as the trajectories pass through parts of
the  instability regions  with large critical exponent $\nu$.
The greatest resonant amplification occurs for a small band of modes
near $k=0$ (see Figure 1) 
since the corresponding
  trajectories lie furthest to the right in the $(q,\, A)$ plane
passing through regions with the highest values of $\nu$. 
Once a trajectory passes below some critical value, $q=q_e$, though, the value
of $\nu$  approaches zero and there is no further amplification.
This means that there is only a finite time for resonant amplification in
the expanding universe, whereas there is no limit in the static case; this is
a key difference that qualitatively changes the nature and robustness of
parametric resonance.  For example, a consequence is that the coupling $g$ 
must exceed a certain minimal value for there to be significant amplification,
as we shall show.  Another prediction  is  that 
increasing $m_x^2 >0$ (or $f(\chi)>0$) suppresses resonant amplification
since it  moves the band of trajectories further
to the left towards the more stable regimes.  

Our Mathieu equation picture can be made quantitative and used
 to estimate
 the energy density
in $X$, $\rho_x(t)$, produced by resonant scattering as a function of
parameters. 
The zero-point contribution of mode $k$ to
the energy density is $\rho_k(t) = \frac 12 \Omega_k(t)$, where
$\Omega_k^2(t)\equiv \omega_k^2(t) + (g^2 \Phi^2/a^3)$. The coupling of
$\Phi$ to $X$ appears as an additional effective mass for the $\chi$ field.
 The production of particles by  parametric resonance adds energy density
\begin{equation}
 \rho_k(t)= \frac 12 \Omega_k(t) ({\rm exp}[2 \int_{t_0}^t dt' \nu_k(t')]-1),
\end{equation}
beginning from time $t_0 = {\cal O}(m_{\phi}^{-1}) = {\cal O}(1)$ when inflation
ends and inflaton oscillations begin; here, $\nu_k(t)$ is the critical exponent
for mode $k$ at time $t$.

Only modes within a small band near $k=0$ contribute significantly to
the final, integrated $\rho_x(t)$.
See Figure 1.
For non-zero comoving wavenumber
$k$, the trajectory $A= 2q + (k/a)^2 + m_x^2 > 2q$ lies to the left of the
$A=2q$ line where
there are  smaller
values of the critical exponent $\nu_k(t)$ on average  and, hence, there is
 less amplification.
As  the trajectory proceeds to smaller $q$,
 the  value of $\nu_k$ averaged over several instability
regions  ($\equiv \bar{\nu}_k(t)$)
increases  as $k/a$ decreases.
To gauge the increase in $\bar{\nu}_k(t)$,
it is useful to consider the family of
curves of the form
$A=2q+\sqrt{bq}$, along each of which the time-averaged 
$\bar{\nu}=\bar{\nu}(b)$ is uniform  ({\it e.g.}, as can be confirmed
using {\it Mathematica 3.0}$_{\em TM}$).
The value of $\bar{\nu}(b)$ increases as $b$ decreases, reaching 
up to $\bar{\nu}(b)\sim 0.1$ for $b=0$.
An important divider is the curve with $\bar{b}={\cal O}(1)$ for
which $\bar{\nu}(\bar{b})
\approx 0.05$.
Modes with  trajectories lying above this curve 
 have very small critical exponents,
$\bar{\nu}_k < \bar{\nu}(\bar{b})\ll 0.1$, and so undergo very 
little amplification. Hence, only the small band of 
modes close
to  $k=0$  with
trajectories lying below this curve
  are significantly amplified. 
 The resonant band  corresponds to
 $0<k^2 \le \bar{k}^2$ where $\bar{k}^2= \sqrt{\bar{b} q_0}-m_x^2$, 
which decreases slightly as $m_x$ increases. (A similar relation has
been introduced based on different arguments in Ref. 6.)  
  Also, $\bar{\nu}_{k=0}$
decreases slightly as $m_x$ increases because the $k=0$ mode is 
shifted to the left in the massive case.
We can approximate
$\int dt' \nu_k(t') \approx \bar{\nu}_{k=0} (t-t_0)$ and obtain
a simple expression for 
the total amplified energy density:
\begin{equation}   \label{rho1}
\rho_{\chi}(t) = \int \frac{d^3 k}{(2 \pi)^3} \rho_k(t) 
\approx
\frac{\bar{k}^3}{6 \pi^2} \Omega_{k=0}(t)({\rm exp}[2 \bar{\nu}_{k=0} (t-t_0)] -1),
\end{equation}
where $\bar{\nu}_{k=0} \approx 0.1$ for $m_x\ll 1$ and 
and $\bar{\nu}_{k=0} \approx 0.075$ for $m_x >1$.

The key parameter,  $\bar{\nu}_{k=0}$, can be
estimated without recourse to  numerical averaging over a trajectory 
once one is familiar with some basic properties of the Mathieu plot.
 From the Mathieu plot,
one can see that $\nu$ varies  between $0$ (its value in the stable regimes)
 and $0.27$ (its maximal value in the unstable regimes).    
For $m_x = 0$, the trajectory moves through stability and instability regimes
such that
the stability regimes are wider than the instability regimes (that is, 
for the range of small $q$'s responsible for  amplification). 
Hence, one can anticipate 
that the average $\nu$ has a value somewhat  less than 
 half   its maximum  ($0.135$).
Indeed, a numerical
integration along the trajectory in the $(q,\,A)$ plane yields the
result
  $\bar{\nu} = 0.12$.

  For the massive case ($m_x \ge 1$), it is evident that $\nu$ has
to be smaller than the massless result, 0.12: the trajectory
lies  further to the left of the $A=2q$ line  where the
critical exponent is lower.
On the other hand, as shown in Figure 1, the trajectory lies to the
right of 
the curve $A = 2q + \sqrt{bq}$.  Along this curve, $\nu$ averaged
over several stability/instability regimes is constant; for $b=1$, say,
the average value of 0.05.  Hence, one can anticipate that the 
average value for $m_x \ge 1$ lies between 0.05 and 0.12.  Numerical
integration shows $\bar{\nu} \sim 0.08$, in agreement with this estimate.

Resonance continues until $q(t) =
q_0/a^3 = q_0 (t_0 /t)^2$ falls below the critical value $q_e$.  
For the massless
case,  $q_e \approx 1/4$ below which  trajectories enter a Mathieu
stability
region that sustains as $q\rightarrow 0$.  For the massive case ($m_x>m_{\phi}=1
$),
resonance sustains until $A=2q_e + \sqrt{\bar{b}q_e} = 2 q_e +m_x^2$, or
$q_e = \frac{1}{\bar{b}} m_x^4$.  
The duration of  the resonance is $t-t_0 = (\sqrt{q_0/q_e}-1)t_0$.
To reach barely efficient amplification ($\rho_x \approx \rho_{\phi}$),
it is necessary that $q_0\gg q_e$.

Putting these details together, we obtain our prediction for general $m_x$
\begin{equation}   \label{rho2}
\rho_{\chi}(t_e) =  \frac{[(\bar{b} q_0)^{1/2}-m_x^2]^{3/2}}{6 \pi
^2} (m_x^2+g^2 \Phi_0^2(t_0/t_e)^2)^{1/2} ({\rm exp}[0.2 \sqrt{q_0/q_e} t_0] -1),
\end{equation}
where $q_0 = g^2 \Phi_0^2/4$, $q_e = {\rm max} \,\{ m_x^4/\bar{b}, 1 \}$, 
$t_e = \sqrt{q_0/q_e}t_0$ and
$t_0 \approx 1$.
For the case of interactions, $f(\chi)>0$, we replace $m_x^2 \rightarrow
m_x^2 +f(\chi)$.

Figure 2 compares our prediction to the results
of our exact numerical code based on
Eqs.~(\ref{eom1})
and (\ref{eom2}). The figure shows an example of efficient
preheating, $\rho_x(t_e) /\rho_{total} = {\cal O}(1)$, and several choices
of parameters which produce
inefficient 
preheating, $\rho_x(t_e) \ll \rho_{\phi} < \rho_{total}$. In both regimes, our 
heuristic picture predicts the general behavior and estimates 
$\rho_x(t_e)/\rho_{total}$
reasonably well.
We include an example with $V(X)=\frac 14 \lambda X^4$, a
case of self-interactions which has been 
discussed qualitatively\cite{Allahverdi97,Riotto97}
and computed numerically.\cite{B.Greene97,Prokopec97}
Here 
we are able to predict quantitatively its effect
simply by modifying 
our expressions for $q_e$ and $\Omega_{k=0}(t)$  
in Eq.~(\ref{rho2}) to take account of $f(\chi)$.

The numerical solutions show fine
and intermediate scale structure which can also be understood in our
Mathieu picture.  Consider, for example, the $\chi$ particle number 
($n_{\chi}$)
 as 
a function of time, as shown in Figure 4(a).
Because $n_{\chi} \propto 1/\Omega_{\chi} \propto 1/\Phi$, there is a 
sharp spike each time $\Phi$ oscillates through zero, the minimum
of its potential.  This is  an artifact of the definition; the 
physically important issue is how the particle number changes between
spikes and  from
 before
to after each spike.  At early times when $q$ is large, one observes
that $n_{\chi}(t)$ is flat between spikes and undergoes a discrete jump
up or down (mostly up) from one side of the spike to the other.
In discussing 
the strong coupled regime, Kofman, {\it et al.}\cite{Kofman97}
 have pointed to this
behavior and argued that the up and down jumps
 become stochastic when $q$ passes
through many stability/instability bands within
one inflaton oscillation and the coupling is large.
However, here  we see the same jump feature 
in the more weakly coupled regime 
even though
 $q$ is not changing as rapidly. And, while there are
occasional downward jumps, the average   behavior is not stochastic; 
they are well fit by an exponential envelope.
This is because the jumps and plateaus 
themselves have nothing to do with stochasticity
or rapid evolution of $q$; they are already features of the static
(constant $q$) Mathieu equation when $q$ is large.  The discrete
jumps are  illustrated
in Figure 4(b) for constant $q=119$.
Because $q$ is constant, all the jumps are upward.  In numerical 
experiments where $q$ varies slowly from a  instability regime to 
a stable regime during an oscillation, we find downward jumps.
Hence, all the small-scale structure in the Figure 4(a) can be 
understood in terms of the static or near static Mathieu equation.

Figure 4(c) shows the solution for the static Mathieu
equation when $q$ is small ($q=0.6$), 
in which case the particle production 
is found to increase continuously between spikes and across spikes.
This behavior compares well with the late time behavior in Figure 4(a)
when $q$ is small.

This behavior can be understood  semi-analytically.
In either the high or 
low $q$ regime, the solution of the static Mathieu equation
is described by a Floquet solution,
\begin{eqnarray}
        \chi = e^{\nu t}f(t).
\end{eqnarray}
That is, there is an exponential envelope multiplying a modulating function,
 $f(t)$.  The difference in the low- versus high-$q$ limit is that the
 modulating
function is continuously increasing
 at low $q$ whereas it has sharp jumps at high-$q$.
The particle number, then, is 
\begin{eqnarray}
        n_{\chi} = e^{2 \nu t} g(t)
\end{eqnarray}
where $g(t)$ is a  function of $f,f',f''$ which has a step-like
behavior for high $q$ and is continuously increasing for low $q$.
All this behavior is reflected in the expanding universe
solutions found numerically.
While the details of the modulating function
are  interesting and it is gratifying that they can be
understood within the Mathieu picture, 
what is important for our purpose of predicting the average 
particle production is the fact that  the
envelope function is exponential and that $\nu$ is predictable by
taking the average over trajectories.
Figure 4 illustrates that the exponential envelope as predicted by
the Mathieu picture  is a good fit
in the static and expanding universe limits in both the high and
low $q$ regime in the weak coupling limit.

In addition to small-scale structure (on the scale of a single
$\phi$ oscillation) in at early times,
the  particle number $n_{\chi}$ in Figure 4 and 
the energy density $\rho_x$ in Figure 2   also 
show intermediate scale structure --- a sequence of
rises and broad plateaus stretching over many oscillations.
These occur because $q$ decelerates as the universe expands
 and
spends several oscillations in each of the last few stability and
instability regions. 
The rises/plateaus  correlate with individual
 stability/instability bands.
During this  period,
 most of the amplification
takes place.

We have argued that the Mathieu picture explains qualitatively all
the features seen in particle production in the inefficient to 
barely efficient regime.   We have also seen from our Figures 
that the  picture
agrees  well quantitatively with numerical results, 
in spite of the fact that it ignores
 stochasticity,
backreaction  and rescattering effects which are included in the 
code.
The code itself is standard. It
 employs a variable time step size Runge-Kutta algorithm.
We approximate the Fourier transform of a field by a discrete lattice
in k-space. Each lattice point corresponds to one of the set of coupled
differential equations. The rescattering is taken
into account by splitting the inflaton field into a zero-mode piece and
a fluctuation piece: $\phi = \phi_0 + \delta\phi$. We treat $\phi_0$ as
 a classical background field, and decompose $\delta\phi$ into k-modes,
with a treatment similar to the of the $\chi$ field.
The good quantitative agreement between the Mathieu picture and 
numerics found support our theoretical argument
that
 backreaction and rescattering are not important
for determining the boundary between inefficient and efficient preheating.
They only become non-negligible for couplings far 
into the efficient preheating regime.

Our analysis predicts that non-negligible resonance occurs
for only  a limited range of $g$.
First, there is a minimum value of
$q_0$ needed to have non-negligible  resonance, which implies a lower
bound on  $g$.  The
values of $m_{\phi}$ and  
 $\Phi_0$
are both fixed by the constraints of having sufficient
e-folds of inflation and sufficiently small density fluctuations.
The only freedom left   to
adjust $q_0 = g^2 \Phi_0^2/4$ is through the coupling $g$.
In order to have $q_0> q_e$,
it is necessary that $g^2 \Phi_0^2 > {\rm max} \, \{
\frac{4}{\bar{b}} m_x^4, \, 1\}$.
However,  there is also  an upper bound on $g$.  The resonance picture 
assumes
 $g^2 \Phi_0^2 \ll M_p^2$, or else the  mass of the $X$-field
is $(m_x^2 + g^2 \Phi^2)^{1/2} > M_p$ and quantum gravity effects
become important.
In our example,
chaotic inflation for $V(\phi)=\frac 12
  m_{\phi}^2 \phi^2$, $\Phi_0 \approx
M_p$ at the end of inflation, so we must  have $g\ll 1$.
 Also,  
 $g<10^{-3}$ is required in order to avoid large
radiative corrections (assuming no supersymmetric cancellations).\cite{Kofman97}

A second prediction is that resonant amplification falls sharply as
$m_x$ increases above the inflaton mass.
  For example, in our model,
 inflation requires  $m_{\phi} \approx 10^{13}$~GeV. A proposed
application of parametric resonance has been to produce $X$-bosons with
mass $10^{14}$~GeV or higher whose decay may produce baryon asymmetry. Our
analysis
suggests that parametric resonance  is not
efficient for $m_x/m_{\phi} \gg 1$ unless $g$ is much larger than 
in the massless case.  

In general, our method can be used to determine
 the range of  $g$ and $m_x$ resulting in
 efficient preheating, as shown in Figure~3.
Inefficient preheating means that parametric resonance is inconsequential
and reheating occurs
predominantly through incoherent decay.
This curve shows that the efficient preheating regime is small -- the minimal value
 of $g$ needed to have efficient preheating
 rises
 sharply with $m_x$, ultimately exceeding $g=1$ (where all approximations
 become invalid) at $m_x \approx 100 $. 
 Hence, parametric resonance cannot be relied
on to produce $X$ particles  many
times more massive than the inflaton field,\cite{Kofman97}
which is problematic for models of high energy baryon asymmetry generation.
Adding interactions with $f(\chi) > 0$ further suppresses  resonance.
In particular, efficient reheating requires a small quartic
coupling, $\lambda \ll g^2$ (see Figure 2).

Although our results have used explicitly
a quadratic inflaton potential,
we have checked that the trajectory picture works well
 a quartic inflaton potential as well, in which
resonant amplification can also be described in terms of stability-instability
regions.\cite{Greene97}
In either case, an issue that remains to be explored is an apparent, classic
chaotic  instability
that occurs in our numerical codes  
for  choices of $g$ and $m_x$ far into the efficient preheating regime of
Figure 3 as backreaction and rescattering become significant.

In summary, we have developed a picture of parametric resonance in an
expanding universe in which resonant $X$-particle production is associated
with passage through a sequence of instability regions of the Mathieu
equation. 
The picture explains qualitatively and quantitatively
why resonance in an expanding universe  is effective for only
a constrained range of couplings, masses and interactions, especially
in the limit of large $m_x$.
In particular, we have shown that efficient preheating
 occurs only if
the  dimensionless couplings satisfy  somewhat unusual  conditions
where the coupling $g$ is large  but 
all other interactions of the inflaton 
and $X$-field are  small.
Unfortunately, this means that   parametric resonance
 in inflationary cosmology is  less generic or robust than was hoped 
for.

We would like to thank  L. Kofman, A. Linde, D. Boyanovsky,  
I. Tkachev and R. Holman  for 
useful discussions in the course of this work. 
This research was supported
by the Department of Energy at Penn, DE-FG02-95ER40893.

\newpage

\newpage
\begin{figure}[h]
\epsfbox{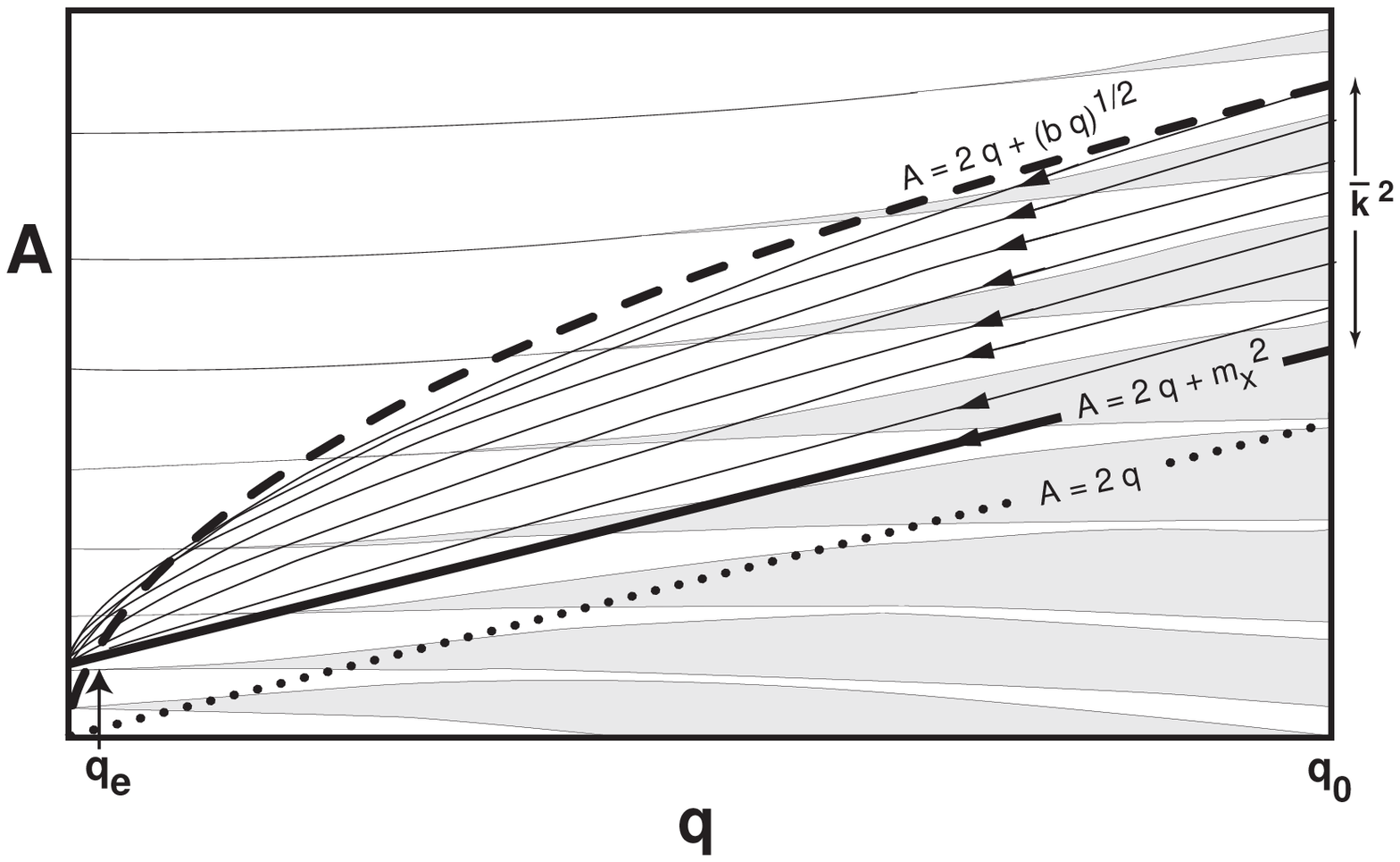}
\caption{ A sketch of the $(q,\, A)$
plane in which  grey (white) strips
represent
instability (stability) regions of the Mathieu equation.
 The dotted line  corresponds to $A=2q$. The 
thick solid line corresponds to $A=2q + m_x^2$, the trajectory
of the $k=0$ mode (assuming $V(X)=0$). 
Also sketched (see thin solid lines with arrows) are   trajectories with $k \le 
\bar{k}$, which dominate the resonant amplification. 
Only trajectories lying below the 
dashed curve, $A=2q +\sqrt{bq}$ with $b= {\cal O}(1)$, 
are significantly amplified; in the text,
we explain how to use this curve
to determine $\bar{k}$ and $q_e$.
}
\end{figure}
\newpage

\begin{figure}[h]
 \epsfxsize=6 in \epsfbox{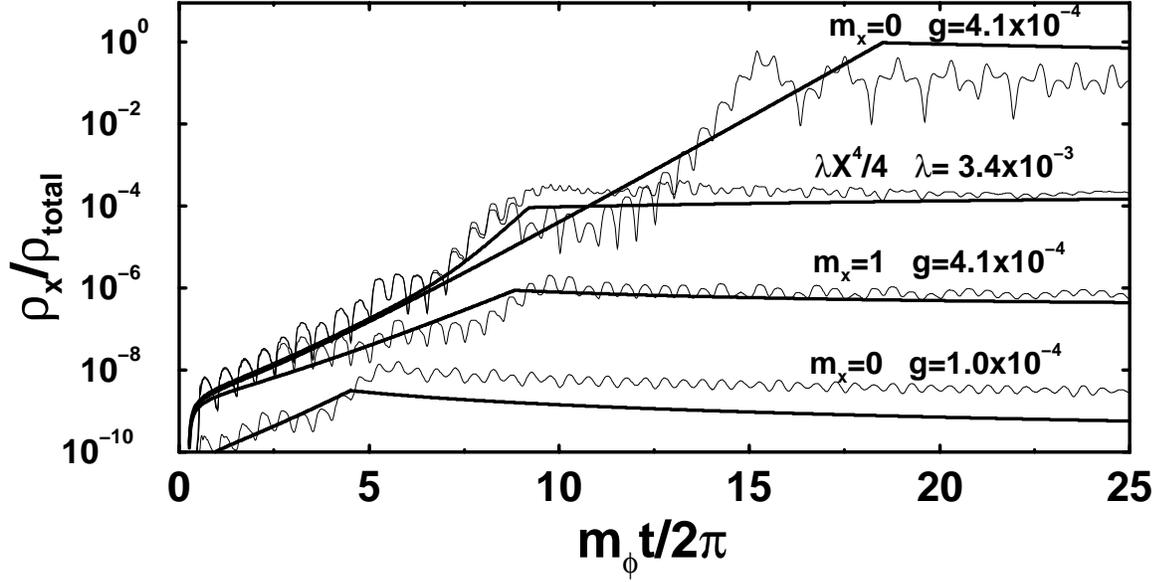}
        \caption[]{ 
Plot of $\rho_x/\rho_{total}$ vs. $m_\phi t/2 \pi$
 comparing numerical results (thin, ragged curves) to
the
heuristic Mathieu equation picture for various models (thick, smooth
curves), where $2 \pi/
m_{\phi}$ is  the  inflaton oscillation period.  The approximation
reliably predicts the time-averaged behavior. The  top pair of curves
corresponds to the minimal $g$ required for the massless case to achieve
efficient preheating, $\rho_x \approx \rho_{\phi}$.
 As predicted by the heuristic picture,
increasing $m_x^2$ or $f(\chi)$ or  decreasing
$g$ greatly suppresses amplification.
}
        \label{}
\end{figure}
\newpage
\begin{figure}[h]
 \epsfxsize=6 in \epsfbox{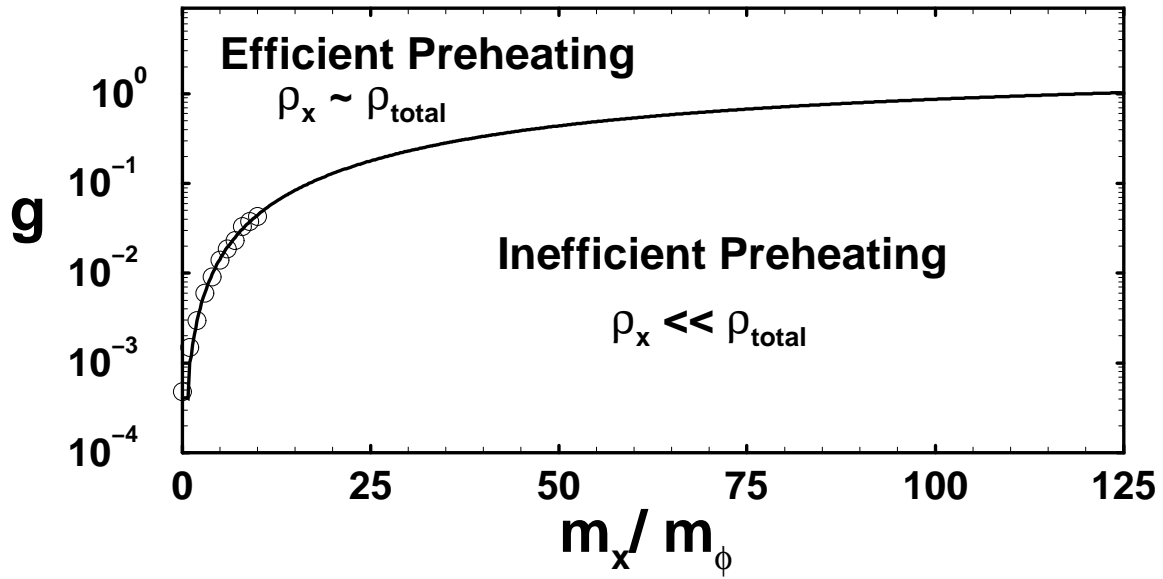}
\caption{A plot of $g$ vs. $m_x/m_{\phi}$ indicating the boundary between
efficient preheating and inefficient
preheating. Numerical results (circles) and
the prediction based on the Mathieu equation (solid curve) are in good 
agreement.
Note the lower bound on $g$ required to have efficient preheating, a 
bound which increases as $m_x$ increases.
  }
\end{figure}

\newpage
\begin{figure}[h]
 \epsfxsize=6 in \epsfbox{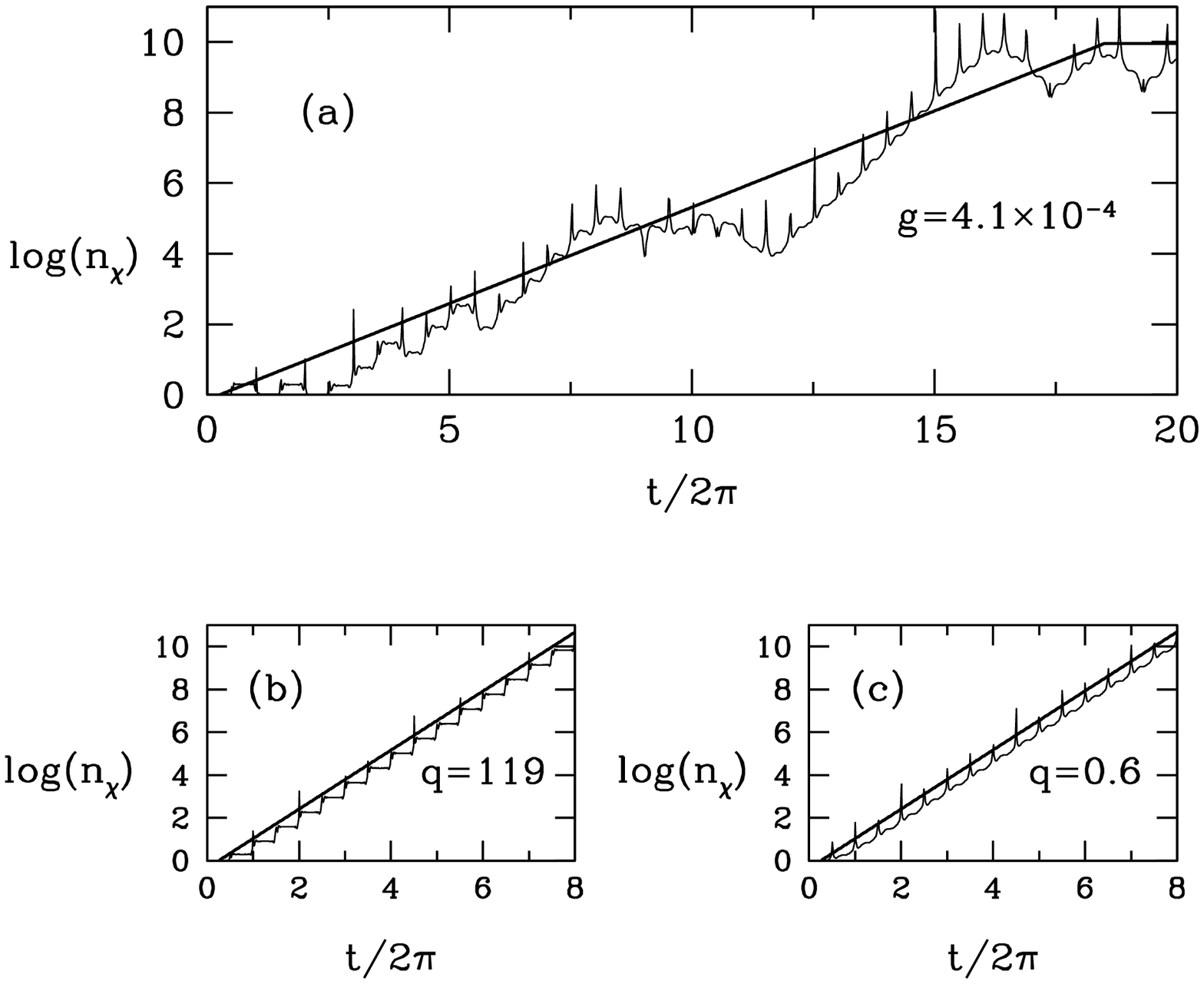}
\caption{
For simplicity, the calculations in all three panels of the figure
are done for a single $k=0$ mode of the $\chi$ field.
Figure (a) shows the number of $\chi$-particles ($n_{\chi}$)
produced as a function 
of time for a typical case of barely efficient preheating.
Each time $\phi$ oscillates through its minimum, there is a sharp
spike; however, particle production is measured by the time-averaged
increase in $n_x$.  Initially, when $q$ is large  
($t/2 \pi \leq 8 $), the
increases in $n_x$ occur in discrete jumps from one side of a spike
to another.
 This behavior is just like the static Mathieu
equation for high $q$, as shown in (b). At later times ($t/2 \pi \geq 8$), 
the growth
is continuous  between spikes and across them, similar to the static
Mathieu equation at small $q$, as shown in (c). 
For $t/2 \pi \geq 12$, there are 
also rises and plateaus stretching many oscillations as $q$ spends
many oscillations in one instability regime.
  In all cases, the  figures show that the
 growth has an exponential envelope with exponent $\nu$ reliably
 predicted by the
 Mathieu picture.
 } 
\end{figure}

\end{document}